\documentclass[%
 aps,
 prb,%
 amsmath,amssymb,
 reprint,
 superscriptaddress,%
 longbibliography
]{revtex4-1}

\usepackage{graphicx}
\usepackage{dcolumn}
\usepackage{bm}
\usepackage{booktabs}
\usepackage{miller}
\usepackage{xcolor}

\begin{document}
\renewcommand\millerskip{}

\title{Irradiation induced modification of the superconducting properties of heavily boron doped diamond}

\author{D.L. Creedon}
 \email{daniel.creedon@unimelb.edu.au}
\affiliation{School of Physics, University of Melbourne, Parkville VIC 3052, Australia}

\author{Y. Jiang}
\affiliation{School of Physics, University of Melbourne, Parkville VIC 3052, Australia}

\author{K. Ganesan}
\affiliation{School of Physics, University of Melbourne, Parkville VIC 3052, Australia}

\author{A. Stacey}
\affiliation{Centre for Quantum Computation \& Communication Technology, School of Physics, University of Melbourne, VIC 3010, Australia}

\author{T. Kageura}
\affiliation{School of Science and Engineering, Waseda University, 3-4-1 Okubo, Shinjuku, Tokyo 169-8555, Japan}

\author{H. Kawarada}
\affiliation{School of Science and Engineering, Waseda University, 3-4-1 Okubo, Shinjuku, Tokyo 169-8555, Japan}

\author{J.C. McCallum}
\affiliation{Centre for Quantum Computation \& Communication Technology, School of Physics, University of Melbourne, VIC 3010, Australia}

\author{B.C. Johnson}
\affiliation{Centre for Quantum Computation \& Communication Technology, School of Physics, University of Melbourne, VIC 3010, Australia}

\author{S. Prawer}
\affiliation{School of Physics, University of Melbourne, Parkville VIC 3052, Australia}

\author{D.N. Jamieson}
\affiliation{Centre for Quantum Computation \& Communication Technology, School of Physics, University of Melbourne, VIC 3010, Australia}

\date{\today}

\begin{abstract}

Diamond, a wide band-gap semiconductor, can be engineered to exhibit superconductivity when doped heavily with boron. The phenomena has been demonstrated in samples grown by chemical vapour deposition where the boron concentration exceeds the critical concentration for the metal-to-insulator transition of \hbox{$n_{MIT} \gtrsim 4\times10^{20}/\text{cm}^3$}. While the threshold carrier concentration for superconductivity is generally well established in the literature, it is unclear how well correlated higher critical temperatures are with increased boron concentration. Previous studies have generally compared several samples grown under different plasma conditions, or on substrates having different crystallographic orientations, in order to vary the incorporation of boron into the lattice. Here, we present a study of a single sample with unchanging boron concentration, and instead modify the charge carrier concentration by introducing compensating defects via high energy ion irradiation. Superconductivity is completely suppressed when the number of defects is sufficient to compensate the hole concentration to below threshold.  Furthermore, we show it is possible to recover the superconductivity by annealing the sample in vacuum to remove the compensating defects.

\end{abstract}

\maketitle

\begin{figure*}[th]
    \centering
    \includegraphics[width=1.0\textwidth]{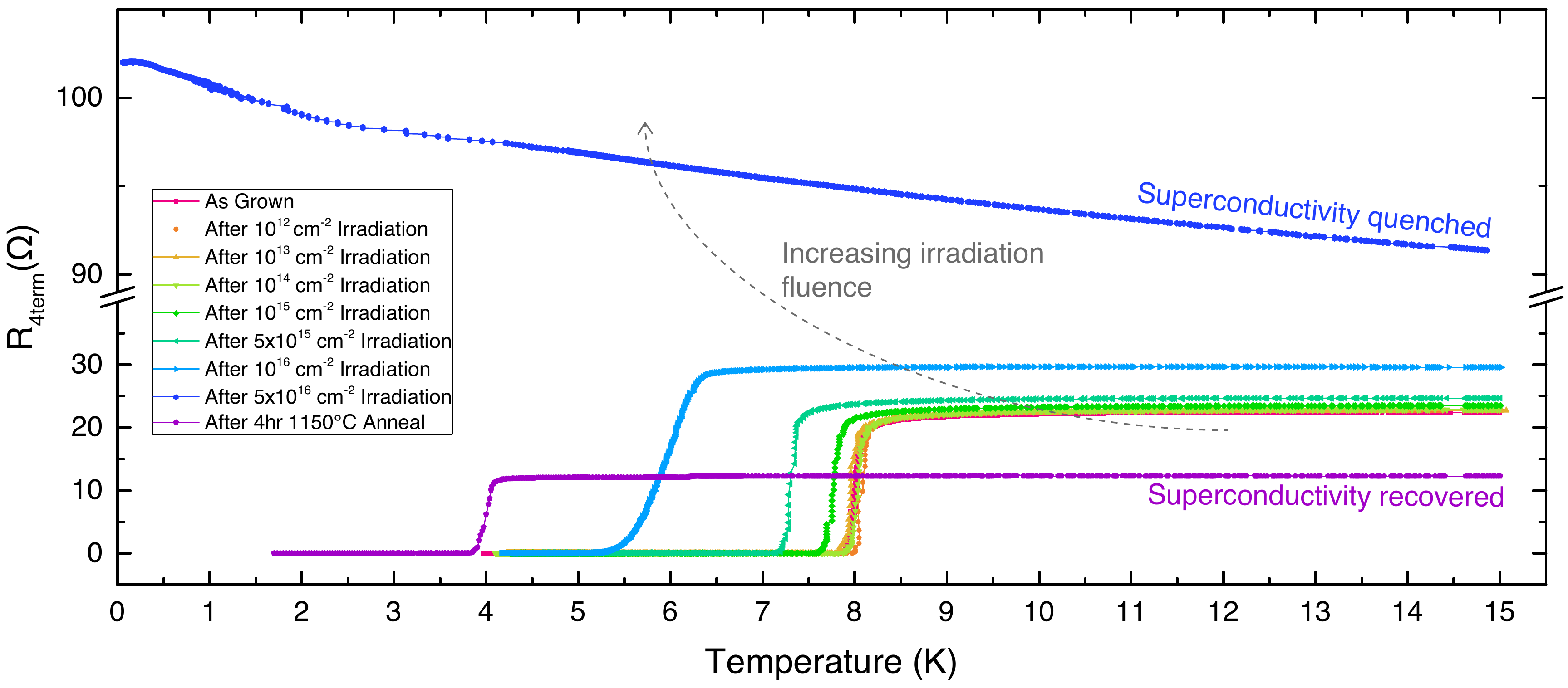}
    \caption{Four terminal resistance measurement performed after each irradiation of the sample. The dashed line  is a guide to the eye showing that the normal-conducting sample resistivity generally increases as a function of irradiation fluence, while the superconducting critical temperature decreases. Note the break in the vertical axis due to the significantly higher $R_{4\text{term}}$ after superconductivity was quenched at the highest fluence of $5\times10^{16}$/cm$^2$.}
    \label{fig:irradiation_study}
\end{figure*}

\begin{table*}[th]
\centering
\caption{Summary of key parameters for the irradiated sample at each stage of measurement, including irradiation fluence, defect concentration in the boron doped layer, measured critical temperature, anneal temperature, and hole density and mobility measured at 300 K.}
\label{table:parameters}
\begin{ruledtabular}
\begin{tabular}{@{}ccccccc@{}}
\toprule
\begin{tabular}[c]{@{}c@{}}Irradiation fluence\\ (He/cm$^2$)\end{tabular} & \begin{tabular}[c]{@{}c@{}}Defect concentration\\ (vacancies/cm$^3$)\end{tabular} & \begin{tabular}[c]{@{}c@{}}Accumulated defect\\ concentration (vac./cm$^3$)\end{tabular} & \begin{tabular}[c]{@{}c@{}}$T_{c(\text{mid})}$ \\ (K)\end{tabular}& \begin{tabular}[c]{@{}c@{}}Annealing\\ temp. ($^{\circ}$C)\end{tabular} & \begin{tabular}[c]{@{}c@{}}Hole density\\ $n_H$ (/cm$^{3}$)\end{tabular} & \begin{tabular}[c]{@{}c@{}}Hole mobility\\ $\mu_H$ (cm$^2$/V.s)\end{tabular} \\ \midrule
\hline
0 & 0 & 0 & 8 & - & 3.85$\times10^{21}$ & 0.998 \\
$1\times10^{12}$ & $5\times10^{16}$ & $5\times10^{16}$ & 8 & - & 3.82$\times10^{21}$ & 1.003 \\
$1\times10^{13}$ & $5\times10^{17}$ & $5.5\times10^{17}$ & 8 & - & 3.97$\times10^{21}$ & 0.967 \\
$1\times10^{14}$ & $5\times10^{18}$ & $5.55\times10^{18}$ & 8 & - & 3.86$\times10^{21}$ & 0.991 \\
$1\times10^{15}$ & $5\times10^{19}$ & $5.56\times10^{19}$ & 7.7 & - & 3.66$\times10^{21}$ & 1.019 \\
$5\times10^{15}$ & $2.5\times10^{20}$ & $3.06\times10^{20}$ & 7.3 & - & 3.50$\times10^{21}$ & 1.015 \\
$1\times10^{16}$ & $5\times10^{20}$ & $8.06\times10^{20}$ & 6 & - & 3.12$\times10^{21}$ & 0.953 \\
$5\times10^{16}$ & $2.5\times10^{21}$ & $3.31\times10^{21}$ & - & - & 1.70$\times10^{21}$ & 0.730 \\
- & - & - & 4 & 1150 & 6.68$\times10^{21}$ & 0.593 \\ \bottomrule
- & - & - & 2.2 & 1250 & 6.85$\times10^{21} $ & 0.220 \\ \bottomrule
\end{tabular}
\end{ruledtabular}
\end{table*}

\section{Introduction}

Superconductivity in boron doped diamond has been a topic of great interest since its discovery in 2004\cite{Ekimov2004}, with the phenomena having since been reported in both single-crystal and polycrystalline samples grown by the High Pressure High Temperature (HPHT) and Chemical Vapor Deposition (CVD) methods.  A wealth of experimental and theoretical investigations of the C:B system exist in the literature, ranging from transport measurements and fundamental materials science of the normal- and superconducting phase of the material, to investigations of novel non-diamond phases of carbon such as superconducting Q-carbon\cite{Bhaumik_2017}. The exceptional physical properties of diamond, coupled with its robust superconducting properties, have motivated numerous applications to devices, including delta doped superlattice structures yielding visible light Bragg mirrors \cite{braggmirror}, diamond-based SQUID devices for magnetometry \cite{diamondsquid}, and superconducting diamond nanostructures \cite{0957-4484-21-19-195303} and nanomechanical resonators\cite{BAUTZE2014100}.

Initial theoretical estimates of the maximum critical temperature ($T_c$) of boron doped diamond were in the range of 140--160 K by virtue of the light atomic mass of the constituent atoms and strong covalent bonds\cite{PhysRevB.74.094520}. By taking into account disorder, local variations of boron concentration in the diamond lattice, and local electron-phonon coupling, the $T_c$ was later predicted to be as high as 55 K for boron concentration between \hbox{10-25 at. \%}, with signatures of superconductivity potentially appearing up to 80 K in regions of high localised order\cite{Moussa2008}. Yet, more than a decade of experimental work has failed to achieve a $T_c$ higher than $\sim$10 K, with some indication of the onset of superconductivity as high as \hbox{25 K} in boron-doped diamond\cite{Okazaki:2015aa}, and 36 K in a boron doped amorphous quenched phase of carbon\cite{Bhaumik_2017}.  More recent theoretical predictions by Moussa \& Cohen\cite{Cohen:2011aa} for phonon-mediated superconductors indicate the possibility that, in the ideal case, diamond could exhibit a maximum critical temperature of $T_c = 290\text{ K}$.

Most experimental data is well described by a standard Bardeen-Cooper-Schrieffer (BCS) model of superconductivity, with phonon-mediated Cooper pairing and a Fermi level shifted inside the valence band \cite{Winzer:2005aa,Yokoya:2005aa}. Boron has one fewer electron than carbon, and is readily incorporated into the diamond lattice with an acceptor level at $E_A=E_V+0.37\text{ eV}$.  An impurity band is formed and broadens as the concentration of isolated substitutional boron increases towards $n_B \gtrsim 4\times10^{20}/\text{cm}^3$, at which point the material undergoes a transition from the semiconducting state to a metallic state, which is widely considered in the literature to coincide with the superconducting transition\cite{Kawano:2010aa,Klein_2007}.  Increasing the boron concentration beyond this metal-to-insulator transition (MIT) point generally corresponds to an increase in the $T_c$, indicating that superconducting boron doped diamond may presently be doped sub-optimally\cite{Takano:2009aa} and a further increase in the active boron concentration may yield higher $T_c$. The push to experimentally realise the prediction of near room-temperature $T_c$ in boron-doped diamond will require the incorporation of substitutional boron at values as high as 25\%. Given that equilibrium processes for growth such as HPHT are limited by the solubility of boron in diamond (approximately 2.8\% \cite{Chen1999}) it is clear that non-equilibrium fabrication methods such as CVD growth and ion implantation must be pursued.  Experimental data suggests that boron concentration is not the only factor in achieving superconductivity, with other authors previously observing significantly different $T_c$ for samples having nominally the same boron concentration - a fact attributed to improved crystallinity\cite{Okazaki:2015aa}. The crystallographic orientation of the substrate used in CVD growth also plays a role, with \hkl<111> oriented diamond always resulting in higher $T_c$ than \hkl<100> oriented diamond with the same boron concentration, which may be due to a higher concentration of electrically inactive boron dimers, or lattice strain increasing the density of states at the Fermi level\cite{Takano:2009aa,Takano:2007aa}.

Recently, Bousquet et al. \cite{Bousquet:2017aa} observed for the first time, in contrast to all previous results, that an intermediate metallic non-superconducting phase exists for boron concentrations above the MIT of $n_B \gtrsim 4\times10^{20}/\text{cm}^3$, and below $n_c^S = 1.1 \pm 0.2 \times10^{21}/\text{cm}^3$, corresponding to the actual critical concentration for the onset of superconductivity. This result was achieved by patterning well-defined Hall bar structures to minimise parasitic currents induced by doping inhomogeneities - i.e. percolation-like current paths through localised areas of higher boron concentration. Whereas that work involved the growth of 26 samples of differing thickness and boron concentration, here we confirm this new result by studying a single sample with fixed boron concentration and systematically compensating boron acceptors with vacancies introduced by light ion irradiation. After introducing vacancies at a density sufficient to compensate the carrier concentration to below the critical level for superconductivity, we confirm that superconductivity is quenched. We then remove the compensating defects by annealing, and show that the hole concentration and superconductivity is recovered.

\section{Experimental Methods}

In order to gain further insight into the mechanism for superconductivity in boron doped diamond, and to determine its robustness to damage and the effect on critical temperature, an irradiation damage study of a superconducting sample was performed.  A sample was grown homoepitaxially on a 3$\times$3 mm Type-Ib \hkl<111> oriented substrate via microwave plasma assisted CVD using the method described in several previous pubications\cite{Kawano:2010aa,Takano:2005aa,Takano:2004aa,Takano:2009aa,Takano:2007aa,Yokoya:2005aa}. The substrate was held at 850$^{\circ}$C with a microwave power of 400 W and chamber pressure of 110 Torr consisting of a dilute gas mixture of methane and trimethylboron (TMB) in hydrogen. The methane concentration was 5\% in hydrogen with a [TMB]/[CH$_4$] ratio of 20,000 ppm, and growth was carried out for 20 minutes resulting in a superconducting BDD cap layer of 200 nm thickness.
The as-grown sample demonstrated $p$-type conduction, with a hole concentration of \hbox{$3.8\times10^{21}$/cm$^{3}$} at ambient temperature, and was measured to have a superconducting critical temperature of $T_c = 8.0$ K and critical field in excess of 9 T. We define $T_c$ as the temperature at which the resistivity of the sample was reduced to 50\% of its normal-state resistivity at \hbox{15 K}.
After growth and initial electrical characterisation, a systematic study was carried out in which the sample was repeatedly subjected to irradiation with a $1$ MeV helium ion beam, followed by electrical characterisation at cryogenic temperatures. The fluence of the ion beam was increased with each irradiation and measurement cycle, ranging between \hbox{$1\times10^{12}$/cm$^{2}$} and \hbox{$5\times10^{16}$/cm$^{2}$}, at which point superconductivity was completely quenched.  In each irradiation, the ion beam was raster scanned over a $4\times4$ mm$^2$ region, ensuring the whole sample was uniformly irradiated. The irradiation was performed with the sample held at room temperature ($295$ K), and under vacuum of $4\times10^{-6}$ Torr, oriented `face on' to the beam at a small misalignment angle of 3$^{\circ}$ to avoid ion channelling.  During the irradiation, the sample was glued and wire-bonded into a ceramic chip carrier. Wire-bonds were made to evaporated Ti/Au electrical contact pads of thickness 100 nm and 20 nm respectively, with the bonds reinforced using a generous thickness of silver paste \hbox{($\sim$300 $\mu$m)}. The contact pads were evaporated in a typical van der Pauw configuration at the edges of the sample, having a diameter more than an order of magnitude smaller than the sample dimensions.

\begin{figure}
    \centering
    \includegraphics[width=1.0\linewidth]{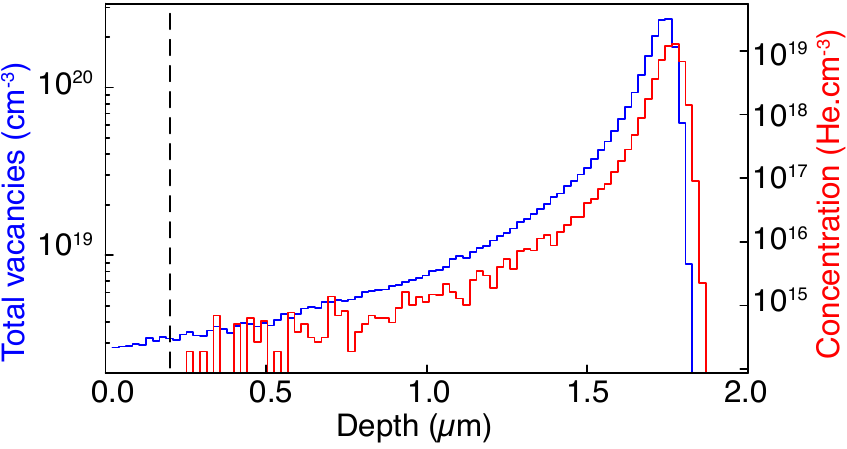}
    \caption{SRIM simulation of 1 MeV He ion irradiation \hbox{($1\times10^{14}\text{ cm}^{-2}$ fluence)} of pure diamond, calculated with a density and displacement energy of of 3.51 g/cm$^3$ and 50 eV, respectively. The 200 nm thickness of the boron doped surface layer is shown with a dashed line to emphasise that the end of range of the ions lies far beyond in the undoped \hkl<111> diamond substrate.  The vacancy profile within the doped layer is relatively uniform, and the end of range of the ions lies approximately 1.5 $\mu$m beyond the interface ($R_P=1.72\mu\text{m}$, $\Delta R_P= 54\text{ nm}$).}
    \label{fig:srim_sim}
\end{figure}

\begin{figure*}[t]
    \centering
    \includegraphics[width=1.0\textwidth]{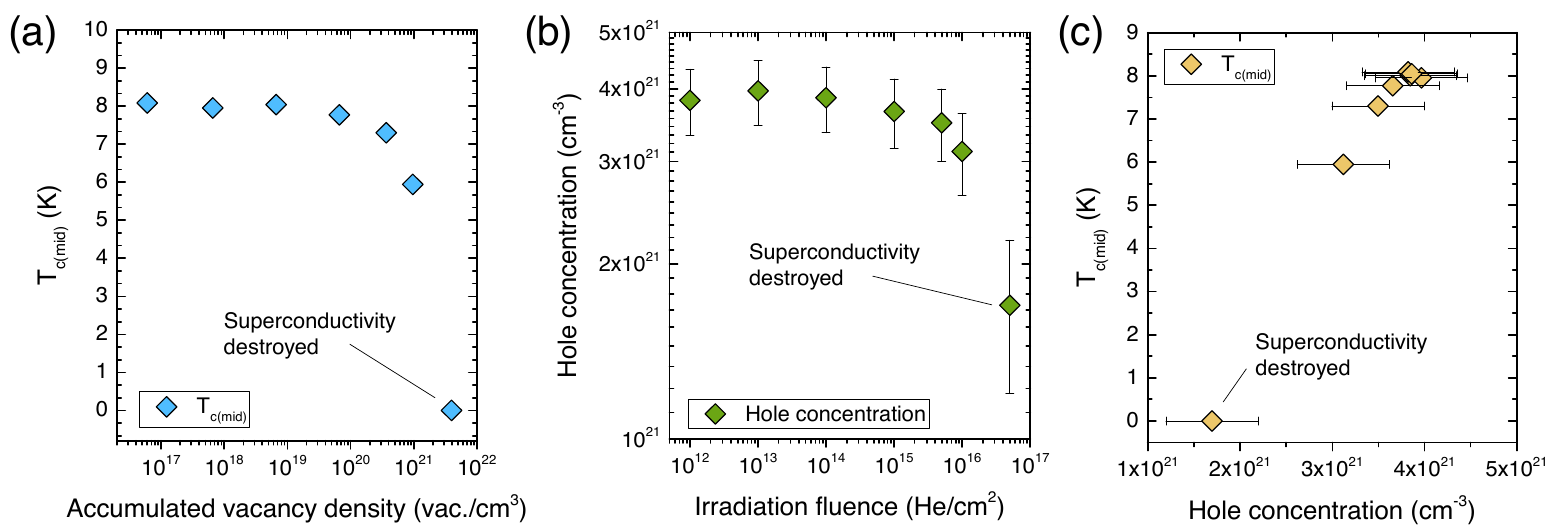}
    \caption{(a) Measured $T_{c(mid)}$ as a function of accumulated vacancy density in the boron doped surface layer, as modelled by SRIM. The $T_c$ is set to zero when superconductivity is completely quenched. Error bars in $T_c$ are smaller than the data point symbols. (b) Bulk carrier (hole) concentration as determined by Hall measurements at 295 K, as a function of irradiation fluence. (c) Measured $T_{c(mid)}$ as a function of hole concentration.}
    \label{fig:tc_vs_vac}
\end{figure*}

The choice of helium as the implant species was made in order to ensure the damage distribution was nominally uniform throughout the boron doped top layer, with the end-of-range of such light ions lying deep inside the substrate, well beyond the doped region.  This was confirmed by modelling the \textit{Stopping and Range of Ions in Matter} (SRIM) software (Figure \ref{fig:srim_sim}).  The heavy damage and amorphization of the lattice from the ions was thus placed deep in the insulating, unmeasured substrate, and only point defects from ion track damage were introduced into the superconducting surface layer. The effect of ion irradiation damage in the contact pad regions is limited to the very surface of the thick reinforcing silver paste layer, effectively shielding the evaporated ohmic contact pads and the superconducting diamond region directly underneath them. The sub-contact regions remaining unirradiated does not affect the electrical measurements, as the path of the current in the van der Pauw measurement is across the bulk irradiated sample region.

Each resultant vacancy acts as a donor which will compensate the hole introduced by a single substitutional boron atom. Thus, without changing the concentration or distribution of boron in the sample, it becomes possible to progressively compensate more and more holes until the superconductivity is quenched. The effect of the strain induced by the buried damage layer (peak depth 1.72 $\mu$m) on the surface boron doped layer is not expected to be significant. Cross-sectional TEM of diamond samples implanted with high energy helium ions shows that the lattice strain is well confined around the amorphized end-of-range of the ions\cite{Fairchild_2012}.

After each irradiation, the sample underwent standard electrical and Hall measurements at room temperature and magnetic fields up to 0.7 T to determine resistivity, mobility, and charge carrier density. The sample was then mounted in a dilution refrigerator and cooled to approximately 50 mK while undergoing four-terminal resistance measurement using standard low noise lock-in amplifier techniques. All measurements were made in a four-terminal van der Pauw configuration using the same pair of contacts for driving current and measuring voltage each time.

\section{Results}

Figure \ref{fig:irradiation_study} shows the four-terminal resistance of the sample below 15 K as a function of irradiation fluence, and Table \ref{table:parameters} shows parameters such as irradiation fluence, defect concentration in the 200 nm superconducting cap layer as modelled by SRIM, superconducting critical temperature, and the hole concentration and mobility determined from electrical measurements. To within the error of the measurement, no change in $T_c$ was seen for an irradiation fluence up to \hbox{$1\times10^{14}$/cm$^{2}$}.  At fluences above this level, three general trends were observed:

\begin{enumerate}
\item The bulk resistivity of the sample (measured cryogenically, above $T_c$) increased with increasing fluence,
\item The superconducting $T_c$ decreased, as well as the charge carrier (hole) concentration $n_H$,
\item The width of the superconducting transition, i.e. the difference between the onset of superconductivity and the offset (zero resistance) was broadened.
\end{enumerate}

The width of the transition in particular suggests an increasingly inhomogeneous material, with the accumulated damage from the ion irradiation creating localised regions of differing defect density and nanoscale order, with a spectrum of critical temperatures. While the sample remained superconducting at 6 K after accumulating a vacancy concentration of \hbox{$8\times10^{20}$/cm$^{3}$}, the final irradiation at a fluence of \hbox{$5\times10^{16}$/cm$^{2}$} led to a large increase in the resistivity of the sample and complete quenching of superconductivity above our base temperature of \hbox{50 mK}. Figure \ref{fig:tc_vs_vac} shows the change in $T_c$ as a function of accumulated vacancy density and hole concentration, and the hole concentration as a function of irradiation fluence.

The efficiency of compensation up to a fluence of \hbox{$1\times10^{16}$/cm$^{2}$}, that is to say the ratio of the measured post-irradiation hole concentration to the expected concentration (pre-irradiation concentration minus the number of vacancies introduced), was very high at 97\%.  The final implant, with fluence \hbox{$5\times10^{16}$/cm$^{2}$}, resulted in lower hole compensation efficiency of only 36\%, but the reason for this is unclear.  In any case, we found that progressively reducing the hole concentration in this superconducting sample also led to a reduction in $T_c$, with the apparent threshold hole concentration for superconductivity lying in the range \hbox{$1.7-3.1\times10^{21}$/cm$^{3}$}.

\begin{figure}[t!]
    \centering
    \includegraphics[width=1.0\linewidth]{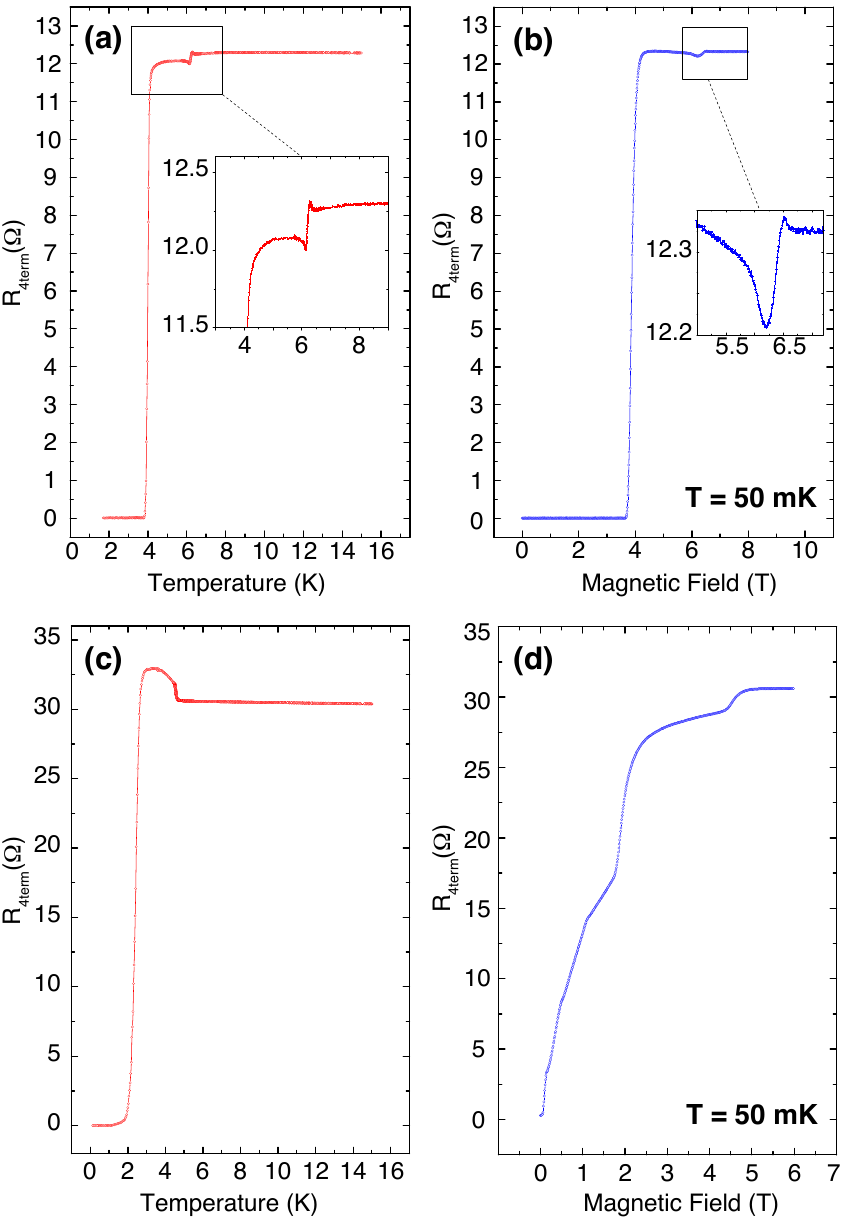}
    \caption{Four-terminal resistance of the sample as a function of (a) temperature, and (b) magnetic field after the recovery of superconductivity by annealing at 1150$^{\circ}$C. The exaggerated vertical scale compared to Figure \ref{fig:irradiation_study} is used to highlight the anomalous transition near 6.5 K. This provides evidence that the material becomes polyphasic in nature after annealing, with an apparent initial superconducting transition at a temperature several degrees higher than the bulk transition of the material. The effect is repeatable, and a similar feature is seen in the critical field measurement of the sample. Subfigures (c-d) show the same measurements after a second anneal at 1250$^{\circ}$C.}
    \label{fig:double_transition}
\end{figure}

It has been shown previously that annealing diamond is an effective method of removing compensating point defects\cite{Kalish:1999aa}. To attempt to remove the point defects introduced by helium ion irradiation, the sample was annealed in vacuum at $1150^{\circ}$C for four hours, with a slow ramp up and ramp down from this temperature of approximately 10 hours each. After annealing, electrical measurements revealed the  successful recovery of superconductivity in the sample, albeit at a lower $T_c$. Figure \ref{fig:double_transition}(a) shows the same data as Figure \ref{fig:irradiation_study} for the sample after annealing at 1150$^{\circ}$C, but with a vertically exaggerated aspect ratio in order to highlight an anomalous preliminary transition before the bulk of the material becomes superconducting.  This behaviour, which was not observed at any point during the irradiation study prior to annealing, suggests that the annealing strategy catalysed the separation of the material into two discrete regions with differing $T_c$. At such a low annealing temperature, boron is not mobile and will not diffuse in diamond, so the redistribution of boron into higher concentration areas is unlikely. We are left to conclude that domains or percolation paths through the material have had compensating defects annealed out with higher efficiency than the bulk, leading to part of the material undergoing a superconducting transition at a higher $T_c$ due to the resultant non-uniform hole density. This echoes the result of Bousquet et al. \cite{Bousquet:2017aa} that highlights the need for well-defined conducting structures to avoid parasitic conduction paths and enable accurate determination of the critical carrier density for superconductivity.\\

After electrical characterisation, the sample was annealed a second time using the same protocol but at a slightly higher temperature of $1250^{\circ}$C. New contact pads were evaporated post-anneal and cryogenic measurement again revealed a seemingly polyphasic material, but with a lower critical temperature than after the first anneal (Figure \ref{fig:double_transition}(c)).  The transition from the normal conducting state to superconductivity is now preceded by an initial rise in resistance, an effect which has been seen previously in boron doped granular diamond\cite{PhysRevLett.110.077001}. In that case, the effect was attributed to a small degree of disorder preventing a direct transition of the material from the metallic state into global superconductivity. Instead, the material is first tuned through a bosonic insulator phase as localised disorder and the associated spatially varying potential cause bosonic `islands' of Cooper pairs to form.  As the temperature is decreased, the bosonic islands grow and begin to percolate, allowing the onset of global superconductivity.  Microscopic disorder has also been directly shown to cause nucleation of superconducting pairing gaps in nanoscale regions above $T_c$ in the high temperature superconductor Bi$_2$Sr$_2$CaCu$_2$O$_{8+\delta}$ \cite{Gomes:2007aa}. These localised islands of superconductivity proliferate as the temperature is lowered, resulting in a spatially distributed collection of domains with varying superconducting $T_c$. The effect could also be indicative of that seen in disordered metals close to the MIT \cite{Osofsky_2001,Osofsky_2002}.

Additional structure is apparent in the measurement of critical field in our sample after the second anneal (Figure \ref{fig:double_transition}(d)), with the field for global superconductivity heavily suppressed from that measured after the first $1150^{\circ}$C anneal (Figure \ref{fig:double_transition}(b)).  As many as six partial transitions can be seen between the metallic state at the highest measured field and global superconductivity near zero field, indicating a highly non-uniform material with a distribution of regions of differing $H_c$.

\section{Discussion}

Whereas previous work has compared CVD boron doped diamond layers of differing thickness, boron concentration, crystallographic orientation, and almost certainly differing local disorder at the nanoscale, the present work has studied a single sample of fixed boron concentration. Irradiation with high energy light ions was used to systematically compensate boron acceptors through the introduction of vacancies, with one single vacancy capable of compensating one single substitutional boron atom.  Thus, the density of states at the Fermi level $E_F$ is reduced, and correspondingly the $T_c$ also decreases.  Our data shows good agreement with this simple picture, with the measured concentration of acceptors (holes from substitutional boron) reduced by the accumulated number of donors (vacancies as modelled by SRIM) with each iteration of the irradiation and measurement process. At the point where the hole concentration is compensated beyond the threshold for superconductivity near the metal-to-insulator transition, superconductivity is quenched as expected.

The recovery of superconductivity in the sample after annealing, however, highlights that the relationship between active boron concentration and $T_c$ is not so straightforward. After annealing at 1150$^{\circ}$C, the measured hole density at ambient temperature is the highest recorded for this sample, approximately 75\% higher than the as-grown sample with  \hbox{$T_c = 8$ K}. Yet despite the increased hole concentration, the critical temperature after annealing is only 4 K, lower than that measured at any stage of irradiation.  After annealing the sample a second time at 1250$^{\circ}$C, the hole density increases further, though not significantly, but the critical temperature drops again to 2.2 K. 

Now we must reconcile the increase in the measured hole concentration and the decrease in $T_c$ post-annealing, given that the amount of boron in the sample is fixed. The annealing process can have only either removed compensating defects in the sample or changed their location or bonding configuration. The increase in carrier concentration can potentially be attributed to the removal of existing compensating defects such as hydrogen in addition to the vacancies introduced during irradiation, effectively activating them electrically. However, if this was the case we would expect an increase in $T_c$, up to a certain limit. Because the only inclusion which contributes carriers to superconductivity in diamond is substitutional boron, there exists a boron concentration where $T_c$ saturates due to boron clustering to form dimers. These dimers do not contribute to the density of states at the Fermi level, thereby reducing the value of $T_c$\cite{Bhaumik_2017}. There is however some evidence to suggest that boron dimers can act as acceptors when incorporated at boron concentrations greater than 3\% (i.e. $5.2\times10^{21}/\text{cm}^3$)\cite{Bourgeois:2006aa}. Theoretical studies of boron-based defects in diamond have shown that all substitutional boron complexes act as acceptors, while complexes containing vacancies or interstitials typically act as donors which act to passivate holes in the valence band\cite{Goss_2006}. Interstitial boron are particularly effective at passivating holes, acting as donors that can compensate up to three holes per boron atom \cite{Moussa2008}. 

Because the annealing process has increased the hole density in our sample, we conclude that the anneal acts to increase the fraction of boron located in substitutional sites without being dimerized or compensated by hydrogen or vacancy defects. This may occur either by removing or reconfiguring the defects into non-boron containing defect complexes, or by enabling interstitial or dimerized boron to become single substitutional boron. Given that boron is not mobile in diamond at the annealing temperatures used, it is more likely that the anneal acts to effectively remove compensating defects. Boron, as a substitutional impurity, could however diffuse though vacancies.  The activation energy for this process is as high as 8.9 eV (the sum of the formation and migration energy), however the presence of existing defects such as the large concentration of vacancies may allow an easy parallel diffusion channel\cite{sung_1995}. In any case, as noted previously\cite{Bourgeois:2006aa} one cannot expect a simple dependence of the free carrier concentration on $T_c$.  The apparent disconnect between the very high hole concentration and very low $T_c$ in our sample post-anneal may be explained by localised disorder as discussed by Moussa et al \cite{Moussa2008}. All valence states in boron doped diamond have strong electron-phonon coupling, and Moussa et al. suggest that the only way in which substitutional disorder can quench superconductivity is if it is strong enough to open a gap in the valence states. After annealing, our sample exhibits clear signs of local disorder such as the broadened superconducting transition, and seemingly multiphasic transitions from the metallic to globally superconducting state including something reminiscent of an intermediate bosonic-insulator phase.  It should also be noted that the hole mobility drops significantly at the point where superconductivity is quenched in our sample, and further again after annealing, suggesting increased scattering of carriers by the defects introduced by the irradiation. 

\section{Conclusion}

Our results indicate a threshold carrier concentration for the onset superconductivity in boron doped diamond of between $n_H = 1.7-3.1\times10^{21}/\text{cm}^{3}$. This result is consistent with the recent findings of Bousquet et al. \cite{Bousquet:2017aa} which show that the threshold boron concentration for superconductivity does not coincide with the metal to insulator transition. Furthermore, we show that the critical temperature of the material has reasonable robustness against irradiation damage, showing no degradation of $T_c$ up to fluences of $1\times10^{14}/\text{cm}^3$, with total destruction of superconductivity at a fluences between $1-5\times10^{16}/\text{cm}^3$.  The fact that superconductivity can be recovered in samples of boron doped diamond which have suffered heavy irradiation damage is promising, and the parameters of the recovered sample suggest that localised disorder in superconducting boron doped diamond has a strong effect on the density of states at the Fermi level and thus the $T_c$

Samples grown by plasma assisted CVD currently demonstrate the highest measured $T_c$ for boron doped diamond, but it remains an open question why superconductivity has not been observed in samples fabricated using high energy implantation of boron into diamond, even when doped at a concentration above the MIT threshold\cite{Heera:2008aa,Beveren:2016aa,Tsubouchi:2012aa}. Whilst the damage and amorphization of the lattice caused by high energy ion implantation are likely explanations for the absence of superconductivity, in-situ dynamic annealing is a promising technique\cite{Tsubouchi:2006aa,Beveren:2016aa} for preventing significant damage and maintaining the important property of high atomic order critical for high temperature superconductivity\cite{Cohen:2011aa}. Despite the use of this technique, boron implanted diamond with carrier concentration above the MIT has still not been made to superconduct. 

 The present result may explain the absence of superconductivity in diamond doped by high energy implantation of boron, which is likely to have a very large number of defects which passivate the boron and significant disorder due to implantation damage. Given the assertion that disorder must be high enough to open a gap in the valence states in order to quench superconductivity, future work should include angle-resolved photoemission spectroscopy (ARPES) measurements of samples irradiated with light ions, or samples fabricated via boron implantation. This direct probe of the band structure may provide further insight into how higher critical temperatures can be achieved in this promising material.
 
 In summary, our work demonstrates for the first time the ability to directly alter the superconducting properties of diamond through ion implantation, an advance which may lead to the ability to selectively write well-defined channels or regions of normal-conducting diamond into superconducting layers, or create the Josephson junction `weak links' required for SQUID devices using a focused ion beam. Such devices engineered fully from diamond are of interest due to the fact that interface and surface state noise is a major source of decoherence for superconducting resonators and quantum devices\cite{oliver_welander_2013}. The ability to create buried superconducting diamond devices through CVD growth and subsequent ion irradiation may suppress or avoid the surface and interface noise suffered by present qubit technology which relies on metal/semiconductor or metal/dielectric heterostructures with complex material interfaces. Finally, the unique combination of thermal and mechanical properties of diamond, combined with it's low absorption in a large transmission window ranging from ultraviolet to infrared wavelengths lend itself well to use in integrated photonic circuits. Single photon detectors based on superconducting NbN nanowires integrated into diamond-based nanophotonic circuits have shown promise\cite{Rath:2015aa,Kahl:16}, but the extremely high critical current of boron doped diamond, and the ability to grow epitaxially on diamond with lower lattice mismatch may provide performance enhancements. Using ion irradiation to define nanowire geometries could allow buried nanowires to be fabricated inside the waveguide of such photonic circuits, rather than evanescently coupled on the surface.

\section*{Acknowledgements}
The authors wish to acknowledge that this research was funded by the Australian Research Council under Grant No. DP150102703, and the United States Air Force Research Laboratory under Agreement No. FA2386-13-1-4055. The U.S. Government is authorised to reproduce and distribute reprints for Governmental purposes notwithstanding any copyright notation thereon. The authors also acknowledge access to ion implantation and/or ion-beam analysis facilities at the ACT node of the Heavy Ion Accelerator Capability funded by the Australian Government under the NCRIS program.

\bibliography{biblio}

\end{document}